\documentstyle[11pt,epsf,fleqn]{article}
\begin{document}
\title{Spontaneous Violation of the CP Symmetry in the Higgs Sector \\
of the Next-to-Minimal Supersymmetric Model}
\author{S.W. Ham$^{(a, \ b)}$, S.K. Oh$^{(a)}$, and H.S. Song$^{(b)}$
\\
{\it $^{(a)}$ Department of Physics, Konkuk University, Seoul 143-701, Korea}
        \\{\it $^{(b)}$ Center for Theoretical Physics,
               Seoul National University} \\
          {\it Seoul 151-742, Korea}
\\
\\
}
\date{}
\maketitle
\thispagestyle{empty}
\begin{abstract}
The spontaneous violation of the CP symmetry in the next-to-minimal
supersymmetric standard Model (NMSSM) is investigated.
It is found that the spontaneous violation of the CP symmetry can
occur in the Higgs sector of the NMSSM for a wide region of
the parameter space of the model, at the 1-loop level where
the radiative corrections due to the top quark and scalar-top quark loops
are found to generate the scalar-pseudoscalar mixings between the
two Higgs doublets of the NMSSM.
In our model, we assume that the masses of the left-handed and
the right-handed scalar-top quarks are not degenerate.
And we investigate our model anaytically: We derive analytical
formulae of the 1-loop mass matrix for the neutral Higgs bosons.
We calculate the upper bound on the lightest neutral Higgs boson mass
under the assumption. It is found to be about 140 GeV for
our choice of parameter values in the presence of
the spontaneous violation of the CP symmetry in the NMSSM.
Thus, the possibility of the spontaneous violation of the CP symmetry
is not completely ruled out in the Higgs sector of the NMSSM if
the masses of the left-handed and the right-handed scalar-top quarks
are not degenerate.
Further, the phenomenology of the $K$-${\bar K}$ mixing within the
context of our model is studied.
The lower bound on CP violating phase in the $K$-${\bar K}$ mixing is
found to increase if either $\tan\beta$ decreases or $A_t$ increases.
\end{abstract}
\vfil

\section{INTRODUCTION}
\hspace*{6.mm}
The violation of the CP symmetry in the weak interactions
has been with us more than several
decades since Christenson {\it et al.} had discovered the CP violating
process in weak $K$ decays [1].
In the standard model (SM) of the electroweak
interactions [2], the violation of the CP symmetry in the weak interactions
is generally explained in terms of the complex phase that exists in the
Cabibbo-Kobayashi-Maskawa (CKM) matrix for the charged weak current [3].
However, if the nature can be described by other theories than the SM,
one may consider other possibilities of the CP violation
than the complex phase in the CKM matrix.
One of the possibilities is that the CP symmetry is spontaneously
violated in the Higgs sector.
If the violation of the CP symmetry be occurred in this way,
it is required that the relevant models should necessarily have at
least two Higgs doublets [4].

Evidently, we know that there are such models which have two Higgs
doublets: The supersymmetric extensions of the standard models.
They are the most strong candidates for the fundamental theory of the
nature beyond the SM, which embrace many important characteristics
of the SM.
As is well known, the supersymmetric extensions of the
standard model need at least two Higgs doublets, in order to give masses
to the up-quark sector and the down-quark sector separately [5].
Therefore, in those supersymmetric models, the CP symmetry
can be violated spontaneously in their Higgs sectors, in principle.

Naturally, many authors have investigated the minimal version of the
supersymmetric standard model (MSSM) [6]. They have found that, at the
tree-level, the vacuum cannot spontaneously violate the CP symmetry.
It is because certain restrictions from the supersymmetry are imposed
on the tree-level Higgs sector of the MSSM to conserve the
CP symmetry.
Thus, radiative corrections are taken into account
in order to see if the CP symmetry be violated spontaneously in
the MSSM at higher level.
A couple of reasonable scenarios have been proposed in which the
radiatively corrected Higgs potential of the MSSM leads actually the
spontaneous violation of the CP symmetry.
An unacceptable side effect of the spontaneous CP violation scenario in
the radiatively corrected Higgs potential of the MSSM is that it
leads to a very light neutral Higgs boson which has already been
ruled out by the Higgs search at LEP1.
Consequently, in order to obtain an experimentally viable spontaneous
violation of the CP symmetry scenario in the SUSY model,
the Higgs sector of the MSSM has to be extended.

Among various non-minimal supersymmetric models the simplest one is the
next-to-minimal supersymmetric standard model (NMSSM), in which a new
neutral Higgs singlet field is introduced to the already-existing
two Higgs doublet fields of the MSSM [7].
The NMSSM is apparently appropriate because, at the tree level,
the upper bound on the lightest neutral Higgs boson mass of
the NMSSM is larger than that of the MSSM.
Moreover, in the NMSSM, the superpotential contains a new dimensionless
coupling coefficient for the cubic interaction between the two Higgs
doublet fields and the Higgs singlet field, which may replace
the parameter $\mu$ in the superpotential of the MSSM by developing
the vacuum expectation value of the singlet Higgs field of the NMSSM.

However, it has been shown that the NMSSM can not produce
the violation of the CP symmetry, at least at the tree level
by Rom${\tilde {\rm a}}$o [8].
It is because the vacuum which are chosen to violate the CP symmetry
is found to have a mode with a negative squared masses
for the neutral Higgs bosons.
Consequently, the inclusion of the higher-order corrections to the NMSSM
Higgs potential have been considered as a next step.

Recently, Babu and Barr [9] have analyzed in the NMSSM at the 1-loop
level to include the radiative corrections if they contribute to the
spontaneous violation of the CP symmetry.
It has been assumed in their analysis that the mass of the left-handed
scalar-top quark is degenerate with that of the right-handed one.
Their analysis show that the mass of the charged Higgs boson does not
change by taking into account these radiative corrections and
its mass is estimated to be smaller than about 110 GeV.
Moreover, there is no relative phase among the vacuum expectation
values in the 1-loop effective potential including these radiative
corrections. In this case, the CP-violating minima are possible only when
$\lambda$ is very small for $\tan \beta$ = 1,
and thus two neutral Higgs bosons become very light.

Independently, Haba, Matsuda, and Tanimoto [10] have investigated the
spontaneous violation of the CP symmetry in the Higgs sector of the NMSSM
using the 1-loop effective potential including radiative
corrections due to top quark, scalar-top quark, bottom quark, and
scalar-bottom quark contributions. Here, the degeneracy in the left-
and the right-handed components have not been assumed.
They have found numerically that the spontaneous violation of the CP
symmetry can occur only in a
very restricted region of the parameter space.
For $\lambda$ = 0.16 and $\tan \beta$ = 1, their numerical calculations
have estimated the upper bound on the lightest neutral Higgs boson
mass to be  about 36 GeV, and the sum of two lightest Higgs boson masses is
around 93 GeV.
Haba {\it et al.} also have calculated the mass of the charged Higgs
boson. It has been found to be dependent crucially on the soft
SUSY breaking scalar-quark mass $m_Q$. It is noticeable that the mass
of the chasrged Higgs boson may be as large as about 721 GeV, which is
considerably different from the case of Babu and Barr.

In this paper we are going a little further into the investigation
of the spontaneous violation of the CP symmetry in the neutral Higgs
sector of the NMSSM.
We also assume no degeneracy between the left-handed and the right-handed
components of the scalar quarks, and we derive the 1-loop corrections
as far as we can analytically, in order to investigate the effects of
parameters of the NMSSM upon the spontaneous violation of the CP symmetry.

At initial stage, our Lagrangian density is assumed too to be invariant
with respect to CP property.
After electroweak symmetry breaking, the CP symmetry is spontaneously
broken in the vacuum state of the potential.
Our assumptions go in parallel with the above two analyses [9, 10]
that the vacuum expectation values of the two Higgs
doublet fields as well as that of the Higgs singlet field are complex
in the Higgs sector of the NMSSM.
Naturally, one of the three CP-violating complex phases can be
eliminated by redefining the Higgs fields.
Thus, the spontaneous violation of the CP symmetry is generated in
terms of the remaining two physical phases.

The radiative corrections to the neutral Higgs boson masses are
investigated by using the effective potential method [11].
The 1-loop effective potential contains the contribution of the
top quark and scalar-top quark contributions.
The difference bewteen the work of Babu and Barr and ours is that we
assume no degeneracy between the left- and the right-handed components
of the scalar quarks, as aforementioned.

Since there is an additional Higgs singlet field in the NMSSM, the mass
matrix for neutral Higgs boson is a 5 $\times$ 5 matrix.
We derive an analytical formula for the neutral Higgs boson mass matrix
at the 1-loop level. This is the difference from the works
of Haba {\it et al.}.
Even though at the tree level the scalar-pseudoscalar mixing
elements between two Higgs doublets are explicitly zero, radiative
corrections generate non-zero values for these elements.
The real symmetric mass matrix can be diagonalized by an orthogonal
matrix and then the NMSSM leads to five neutral Higgs bosons.
In the spontaneous violation of the CP symmetry scenario, the
theoretical upper bound on the lightest neutral Higgs boson mass of
the NMSSM is calculated in the
context of the effective potential formalism.

Then, we apply our result to the CP violations in $K$-${\bar K}$ mixing
numerically.
Within the parameter space of the NMSSM that we consider,
the CP violation in $K$-${\bar K}$ mixing are studied within the
framework of the NMSSM.
We find that even though the scalar Higgs bosons have not been discovered
yet at LEP2, the possibility of the spontaneous violation of the CP symmetry
is not completely ruled out in the Higgs sector of the NMSSM.

This paper is organized as follows.
In the next section we breifly describe the Higgs sector of the NMSSM
within the scenario of the spontaneous violation of the CP symmetry
in the Higgs sector of the NMSSM at the tree level, and then at the 1-loop
level in the third section.
The radiative corrections to the tree level Higgs potential are then taken
into account to generate a viable CP-violating vacuum in the Higgs sector.
The mixings between two Higgs doublets can occur through the
radiative corrections due to the contributions from the top quark and the
scalar-top quark contributions.
The exact analytical expressions of the elements of the
neutral Higgs boson mass matrix are derived in our scenario.
At the 1-loop level, the parameter space of the NMSSM, and hence
masses of the scalar Higgs bosons are constrained by
the negative experimental results from LEP1.
The upper bound on the lightest neutral Higgs boson mass is
calculated to be about 140 GeV for our choice parameter values in
the presence of the spontaneous violation of the CP symmetry in the NMSSM.
In the fourth section, we apply our model into the case of the
CP violation in the $K$-${\bar K}$ mixing within the context of the
spontaneous violation scenario in the NMSSM Higgs sector.
The conclusions are given in the last section.
\section{NEUTRAL HIGGS SECTOR AT TREE LEVEL}
\hspace*{6.mm}
Here in this section, we briefly review the concept of the spontaneous
violation of the CP symmetry in the NMSSM at the tree level.
As is well known, the Higgs sector of the NMSSM consists of two Higgs
doublet superfields $H_1^T = (H_1^0, \ H_1^-)$
and $H_2^T = (H_2^+, \ H_2^0)$,
plus a Higgs singlet superfield $N$.
The superpotential of the NMSSM contains only terms with dimensionless
couplings.
Ignoring all quark and lepton Yukawa couplings except for that of the top
quark, the relevant part for the superpotential can be written as
\[
W = h_t Q H_2 t_R^c + \lambda N H_1 H_2 - {k \over 3} N^3 \ ,
\]
where we denote for simplicity
that $H_1 H_2 = H_1\epsilon H_2 =H_1^0 H_2^0 - H_1^- H_2^+$.
The superfield $Q^T = (t_L, \ b_L)$ consists of the left-handed quarks of
the third generation and the superfield $t_R^c$ denotes the charge
conjugate of the right-handed top quark.

In the above superpotential, a global U(1) Peccei-Quinn
symmetry is explicitly broken by the presence of the cubic term in $N$.
Not for the term, i.e., if $k= 0$, the Peccei-Qiunn symmetry would persist
in the NMSSM superpotential, and the tree-level Higgs potential would
lead to a massless pseudoscalar Higgs boson, or the pseudo-Goldstone boson,
after electroweak gauge symmetry breaking.

The tree level Higgs potential in the NMSSM can be written in terms of
$F$-terms, $D$-terms, and the soft SUSY breaking terms as
\[
    V_{\rm tree} = V_F + V_D + V_{\rm soft}
\]
where
\begin{eqnarray}
    V_F &=& |\lambda|^2[(|H_1|^2+|H_2|^2)|N|^2+|H_1 H_2|^2]
            + |k|^2|N|^4 -(\lambda k^*H_1H_2N^{*2}+ {\rm H.c.}) \ , \cr
    V_D &=& {1 \over 8} (g_1^2 + g_2^2) (|H_2|^2-|H_1|^2)^2 \ , \cr
    V_{\rm soft} &=& m_{H_1}^2|H_1|^2+m_{H_2}^2|H_2|^2+m_N^2|N|^2
     - (\lambda A_\lambda H_1 H_2 N + {1\over 3} k A_k N^3+ {\rm H.c.})\ ,
\end{eqnarray}
with $g_1$ and $g_2$ being the U(1) and SU(2) gauge coupling constants,
respectively.

We assume that there is no explicit violation of the CP symmetry
in the Higgs sector [12]. Therefore, the parameters $\lambda$, $k$,
$A_{\lambda}$, and $A_k$ are assumed to be all real.
On the contrary, we assume that the violation of the CP symmetry can
occur spontaneously through the existence of the complex phases in
the vacuum expectation values of the three neutral Higgs fields:
\begin{eqnarray}
<0| H_1^0 |0> & = & v_1 e^{i \varphi_1}  \ , \cr
<0| H_2^0 |0> & = & v_2 e^{i \varphi_2}  \ , \cr
<0| N |0> & = & x e^{i \varphi_3}  \ ,
\end{eqnarray}
where $v_1$, $v_2$, and $x$ are assumed to be positive, and
three non-trivial phases are introduced into the vacuum expectation
values. As usual, $\tan \beta$ is defined as $v_2/v_1$, and
the top quark mass is generated by $v_2$ of the Higgs doublet $H_2$ as
$m_t = h_t v_2$.
One of the three complex phases can be eliminated by redefining the
Higgs fields.
The remaining two physical phases may be chosen as [10]
\begin{eqnarray}
\theta & = & \varphi_1 + \varphi_2 + \varphi_3 \ , \cr
\delta & = & 3 \varphi_3 \ .
\end{eqnarray}

In the spontaneous violation scenario of the CP symmetry, the vacuum is
defined as the stationary point with respect to the two CP violating
phases $\theta$ and $\delta$. This stationary point or the minimum
point is the CP violating vacuum.
Thus, the CP violating vacuum with respect to these phases  satisfies
two minimum equations
\begin{eqnarray}
        & & {\partial <V_{\rm tree} (v_1,v_2,v_3,\theta,\delta)>
                \over \partial\delta} =0 \ , \cr
        & & {\partial <V_{\rm tree} (v_1,v_2,v_3,\theta,\delta)>
                \over \partial\theta} =0 \ .
\end{eqnarray}
Furthermore, one can use the three minimum conditions for $v_1$, $v_2$, and
$v_3$ to eliminate the soft supersymmetry-breaking
masses $m_{H_1}^2$, $m_{H_2}^2$, and $m_N^2$ in the Higgs potential.

Now, in terms of those vacuum expectation values, the three neutral
Higgs fields can be rewritten by shifting them around the CP
violating vacuum. They have three scalar components and three
pseudoscalar components: One of the mass eigenstates of the
three pseudoscalar components is a massless Goldstone mode.
This Goldstone mode can be gauged away by a unitary gauge tansformation,
and we are left with five components for the neutral Higgs fields.
Consequently, the three neutral Higgs fields may be expressed in terms of
the five components as
\begin{eqnarray}
        H_1^0 & = & e^{i \varphi_1} \left \{v_1 + {1 \over \sqrt{2}}
        (S_1 + i \sin \beta P) \right \}  \ , \cr
        H_2^0 & = & e^{i \varphi_2} \left \{v_2 + {1 \over \sqrt{2}}
        (S_2 + i \cos \beta P) \right \}  \ , \cr
        N & = & e^{i \varphi_3} \left \{x + {1 \over \sqrt{2}}
        (X + i Y) \right \} \ ,
\end{eqnarray}
where $S_1$, $S_2$, and $X$ are the scalar components, and $P$ and $Y$ are
the pseudoscalar components.

The mass matrix for the five neutral Higgs bosons is obtained by the
second derivatives of the Higgs potential with respect to the
corresponding Higgs fields evaluated at the CP violating vacuum as
a symmetric 5 $\times$ 5 matrix:
\[
        M_{ij}^2 = M_{ji}^2
        = \left[ \begin{array}{cc}
        M_{S_1, \ S_2, \ X}^{S_1, \ S_2 \ X}
        & M_{S_1, \ S_2, \ X}^{A, \ Y} \cr
                                     &                              \cr
        (M_{S_1, \ S_2, \ X}^{A, \ Y})^T
        & M_{A, \ Y}^{A, Y}
        \end{array} \right]  \ ,
\]
where, respectively, the upper-left 3 $\times$ 3 submatrix and
the lower-right 2 $\times$ 2 submatrix correspond to the scalar part
and pseudoscalar part.
The upper-right 3 $\times$ 2 as well as the lower-left submatrices
correspond to the scalar-pseudoscalar mixing part in the nutral Higgs
boson matrix.
As aforementioned, if the CP symmetry is conserved in the Higgs sector,
the submatrix for the scalar-pseudoscalar mixing would not exist
in the neutral Higgs boson mass matrix.

We obtain the tree-level elements for the mass matrix of the
five neutral (scalar and pseudoscalar) Higgs bosons explicitly as
\begin{eqnarray}
M_{11}^2 & = & (m_Z \cos \beta)^2
+ \lambda x (A_{\lambda} \cos \theta + kx \cos (\theta - \delta))
\tan \beta  \ , \cr
M_{22}^2 & = & (m_Z \sin \beta)^2
+ \lambda x (A_{\lambda} \cos \theta + kx \cos (\theta - \delta))
\cot \beta  \ , \cr
M_{33}^2 & = & (2 k x)^2 - k x A_k \cos \delta
+ {\lambda \over 2x} v^2 A_{\lambda} \sin 2 \beta \cos \theta
             \ , \cr
M_{44}^2 & = & {2 \lambda x (A_{\lambda} \cos \theta
+ kx \cos (\theta - \delta)) \over \sin 2 \beta}  \ , \cr
M_{55}^2 & = & {\lambda v^2 \over 2x} A_{\lambda} \sin 2 \beta \cos \theta
+ 3 k x A_k \cos \delta + 2 \lambda k v^2 \sin 2 \beta
\cos (\theta - \delta)             \ , \cr
M_{12}^2 & = & (\lambda^2 v^2 - {1 \over 2} m_Z^2) \sin 2 \beta
- \lambda x (A_{\lambda} \cos \theta + k x \cos (\theta - \delta))
 \ , \cr
M_{13}^2 & = & 2 \lambda^2 x v \cos \beta - \lambda v \sin \beta (A_{\lambda}
\cos \theta + 2 k x \cos (\theta - \delta)) \ , \cr
M_{14}^2 & = & 0 \ , \cr
M_{15}^2 & = &\mbox{} - 3 \lambda k v x \sin \beta
\sin (\theta - \delta) \ , \cr
M_{23}^2 & = & 2 \lambda^2 x v \sin \beta
- \lambda v \cos \beta (A_{\lambda} \cos \theta + 2 k x
\cos (\theta - \delta)) \ , \cr
M_{24}^2 & = & 0 \ , \cr
M_{25}^2 & = &\mbox{} - 3 \lambda k v x \cos \beta
\sin (\theta - \delta) \ , \cr
M_{34}^2 & = &\mbox{} \lambda k v x \sin (\theta - \delta) \ , \cr
M_{35}^2 & = & 2 \mbox{} \lambda k v^2 \sin 2 \beta
\sin (\theta - \delta) \ , \cr
M_{45}^2 & = &\mbox{} \lambda v (A_{\lambda} \cos \theta
- 2 k x \cos (\theta - \delta)) \ .
\end{eqnarray}
Here, one should note that $M^2_{14} = M^2_{24} = 0$ implies that
there are no scalar-pseudoscalar mixings between the two Higgs
doublets (between $H_1$ and $H_2$) in the Higgs sector.
That is, the spontaneous violation of the CP symmetry is induced by
the presence of the scalar-pseudoscalar mixing terms between
among Higgs doublets and the Higgs singlet (between $H_1H_2$ and $N$) and
the cubic term of the singlet itself (among $N^3$) at the tree level.

It can be seen that all the scalar-pseudoscalar mixings vanish if
$\theta = \delta$; in this case, at the tree level, the spontaneous
violation of the CP symmetry can not occur in the Higgs sector.
Then, the five neutral Higgs bosons are decomposed as
three scalar Higgs bosons and two pseudoscalar Higgs bosons, and
there is no mixing between the scalar and the pseudoscalar Higgs bosons:
The $5\times5$ mass matrix is decomposed as
a $3\times3$ and a $2\times2$ submatrices.
Therefore, in the NMSSM, the spontaneous violation of the CP symmetry
can be realized by the presence of two phases $\theta$ and $\delta$.

However, assuming that the two CP violation phases are not equal,
the scalar-pseudoscalar mixing may happen inevitably in the
Higgs sector, but it has been observed that in large areas of
the parameter space of the NMSSM the spontaneous violation does not
occur at the tree level, because one can always find a mode with
a negative mass squared at the CP violating vacuum of the
Higgs potential [8].
\section{NEUTRAL HIGGS SECTOR AT 1-LOOP LEVEL}
\hspace*{6.mm}
Now, we turn to the 1-loop level.
Since the radiative corrections due to the top quark and scalar-top quark
contributions give significant contributions to the tree level Higgs boson
masses, we include these contributions in order to see their effects
on the spontaneous violation of the CP symmetry.
The full Higgs potential is composed of the tree level part and the
1-loop level part written as
\[
        V = V_{\rm tree} + V_{\rm 1-loop} \ .
\]
According to the Coleman-Weinberg mechanism [14],
the 1-loop effective potential including the contributions of
the top quark and scalar-top quark loops is given by
\[
   V_{\rm 1-loop} = \frac{3}{32\pi^2} \left \{ {\cal M}_{\tilde{t_i}}^4
   \left (\log {{\cal M}_{\tilde{t_i}}^2 \over \Lambda^2} - {3\over 2} \right )
   - 2 {\cal M}_t^4 \left (\log {{\cal M}_t^2 \over \Lambda^2}
   - {3\over 2} \right ) \right \}  \ ,
\]
where ${\cal M}_{{\tilde t}_i}^2 (i = 1, \ 2)$ are the field dependent
scalar-top quark masses, ${\cal M}_t^2$ is the field dependent top quark
mass, and $\Lambda$ is the renormalization scale.
If the top quark mass and the scalar-top quark mass are identical,
then there is no net contribution from the radiative corrections
to the tree level Higgs potential.

After spontaneous electroweak symmetry breaking, the masses of
the left-handed and the right-handed scalar-top quarks
are obtained from the 2 $\times$ 2 mass matrix for them.
They are
\begin{eqnarray}
        m_{\tilde {t}_1, \ \tilde {t}_2}^2 & = &
        m_t^2 + {1\over 2}(m_Q^2 + m_T^2)
        +{1 \over 4} m_Z^2 \cos 2 \beta     \cr
        & &\mbox{}\mp \left [
        \left \{ {1\over 2} (m_Q^2 -m_T^2)
        + \left ({2 \over 3} m_W^2 - {5 \over 12} m_Z^2 \right) \cos 2 \beta
        \right \}^2 \right. \cr
        & &\left. \mbox{} + m_t^2 \left (A_t^2 + \lambda^2 x^2 \cot^2 \beta
        + 2 A_t \lambda x \cot \beta \cos \theta \right) \rule{0mm}{6mm}
        \right ]^{1 \over 2} \ ,
\end{eqnarray}
where $m_Z^2 = (g_1^2 + g_2^2) v^2/2$ and $m_W^2 = g_2^2 v^2/2$ for
$v = \sqrt{v_1^2 + v_2^2}$ = 175 GeV.

In the above equation, the scalar-top quark masses contain the mixing term
between the left-handed and the right-handed scalar-top quarks
as well as the terms proportional to the gauge couplings.
If the contributions coming from the $D$-terms in the scalar-top quark mass
matrix are neglected, the scalar-top quark masses possess a symmetry under
interchange of $m_Q$ and $m_T$.
Moreover, one may notice that the scalar-top quark masses possess
one CP violation phase $\theta$. If the left-handed and the right-handed
scalar-top quarks are degenerate in mass, the mixing term
between the scalar-top quarks would vanish and there would be no
CP phase in the 1-loop effective potential.
The radiative corrections due to the top quark and scalar-top quark
contributions do not shift the CP violating vacuum along $\delta$ direction.
On the contrary, a shift of the CP violating vacuum along $\theta$ direction
is induced by these radiative corrections because the scalar-top quark
masses depend on $\theta$.

Now, the two minimum equations
\begin{eqnarray*}
        & & {\partial <V(v_1,v_2,v_3,\theta,\delta)>
                \over \partial\delta} =0 \ , \cr
        & & {\partial <V(v_1,v_2,v_3,\theta,\delta)>
                \over \partial\theta} =0 \ .
\end{eqnarray*}
at the 1-loop level yield
\begin{eqnarray}
        \tan \delta & = & {3 \lambda v^2 \sin 2 \beta \sin \theta
        \over 3 \lambda v^2 \sin 2 \beta \cos \theta + 2 A_k x} \ , \cr\cr
        k & = & - {16 \pi^2 v^2 \sin^2 \beta A_{\lambda} \sin \theta
        - 3 m_t^2 A_t \sin \theta f(m_{\tilde{t}_1}^2, \ m_{\tilde{t}_2}^2)
        \over 16 \pi^2 v^2 x \sin^2 \beta \sin (\theta - \delta)}  \ ,
\end{eqnarray}
where the function $f$ arising from radiative corrections is defined by
\[
f(m_1^2, \ m_2^2) = {1 \over (m_2^2-m_1^2)}
\left[  m_1^2 \log {m_1^2 \over \Lambda^2} -m_2^2
\log {m_2^2 \over \Lambda^2} \right] + 1  \ .
\]
These are the conditions that are satisfied by the 1-loop CP violating
vacuum.

The full mass matrix at 1-loop level for the
five neutral Higgs bosons is given by
\[
         M^2 = M_{ij}^2 + \delta M_{ij}^2
\]
where $\delta M_{ij}^2 = \delta M_{ji}^2$ denotes the 1-loop level
Higgs boson mass matrix elements, and $M_{ij}^2$ is the tree-level
mass matrix obtained in the previous section.
We calculate the exact analytical formulae for the elements of the
neutral Higgs boson mass matrix at the 1-loop level. Our results
are given by the complicated but exact expressions:
\begin{eqnarray}
\delta M_{11}^2 & = & {3 \over 8 \pi^2}
\left \{ {m_t^2 \lambda x \Delta_1 \over v \sin \beta}
+ {\cos \beta \Delta \over 2 v} \right \}^2
{g(m_{\tilde{t}_1}^2, \ m_{\tilde{t}_2}^2) \over
(m_{\tilde{t}_2}^2 - m_{\tilde{t}_1}^2)^2}
+ {3 m_Z^4 \cos^2 \beta \over 128 \pi^2 v^2}
\log {m_{\tilde{t}_2}^2  m_{\tilde{t}_1}^2 \over \Lambda^4} \cr
& & \cr
& &\mbox{}+ {3 \over 16 \pi^2 v^2}
\left \{
{2 m_t^2 \lambda x A_t \cos \theta \over \sin 2 \beta}
- \left( {4 m_W^2 \over 3} - {5 m_Z^2 \over 6} \right)^2 \cos^2 \beta
\right \}
f(m_{\tilde{t}_1}^2, \ m_{\tilde{t}_2}^2)   \cr
& & \cr
& &\mbox{} + {3 m_Z^2 \cos \beta \over 16 \pi^2 v}
\left \{ {m_t^2 \lambda x \Delta_1 \over v \sin \beta}
+ {\cos \beta \Delta \over 2 v} \right \}
{\displaystyle \log (m_{\tilde{t}_2}^2 / m_{\tilde{t}_1}^2)
 \over (m_{\tilde{t}_2}^2 - m_{\tilde{t}_1}^2)}   \ , \cr
& & \cr
\delta M_{22}^2 & = & {3 \over 8 \pi^2}
\left \{ {m_t^2 A_t \Delta_2 \over v \sin \beta}
- {\sin \beta \Delta \over 2 v} \right \}^2
{g(m_{\tilde{t}_1}^2, \ m_{\tilde{t}_2}^2) \over
(m_{\tilde{t}_2}^2 - m_{\tilde{t}_1}^2)^2}
- {3 m_t^4 \over 4 \pi^2 v^2 \sin^2 \beta} \log{m_t^2 \over \Lambda^2} \cr
& & \cr
& &\mbox{}+ {3 \over 16 \pi^2 v^2}
\left \{
{m_t^2 \lambda x A_t \cot \beta \cos \theta \over \sin^2 \beta}
- \left( {4 m_W^2 \over 3} - {5 m_Z^2 \over 6} \right)^2 \sin^2 \beta
\right \}
f(m_{\tilde{t}_1}^2, \ m_{\tilde{t}_2}^2)   \cr
& & \cr
& &\mbox{} + {3 \sin \beta \over 16 \pi^2 v}
\left ({4 m_t^2 \over \sin^2 \beta} - m_Z^2 \right)
\left \{ {m_t^2 A_t \Delta_2
\over v \sin \beta} - {\sin \beta \Delta \over 2 v} \right \}
{\displaystyle \log (m_{\tilde{t}_2}^2 / m_{\tilde{t}_1}^2)
 \over (m_{\tilde{t}_2}^2 - m_{\tilde{t}_1}^2)} \cr
& & \cr
& &\mbox{} + {3 \over 32 \pi^2 v^2}
\left ({2 m_t^2 \over \sin \beta} - {m_Z^2 \sin \beta \over 2} \right)^2
\log {m_{\tilde{t}_2}^2  m_{\tilde{t}_1}^2 \over \Lambda^4} \ , \cr
& & \cr
\delta M_{33}^2 & = & {3 m_t^4 \lambda^2 \cot^2 \beta \over 8 \pi^2}
\left ( {\Delta_1 \over m_{\tilde{t}_2}^2
- m_{\tilde{t}_1}^2 } \right )^2 g(m_{\tilde{t}_1}^2, \ m_{\tilde{t}_2}^2) \cr
& &\mbox{} + {3 m_t^2 \lambda A_t \cot \beta \cos \theta \over 16 \pi^2 x}
f(m_{\tilde{t}_1}^2, \ m_{\tilde{t}_2}^2)  \ , \cr
& & \cr
\delta M_{44}^2 & = & {3 m_t^4 \lambda^2 x^2 A_t^2 \sin^2 \theta
\over 8 \pi^2 v^2 \sin^4 \beta}
{g(m_{\tilde{t}_1}^2, \ m_{\tilde{t}_2}^2) \over (m_{\tilde{t}_2}^2
- m_{\tilde{t}_1}^2 )^2}
+ {3 m_t^2 \lambda x A_t \cos \theta \over 16 \pi^2 v^2 \sin^3 \beta \cos \beta}
f(m_{\tilde{t}_1}^2, \ m_{\tilde{t}_2}^2)  \ , \cr
& & \cr
\delta M_{55}^2 & = & {3 m_t^4 \lambda^2 A_t^2 \cot^2 \beta \sin^2 \theta
\over 8 \pi^2}
{g(m_{\tilde{t}_1}^2, \ m_{\tilde{t}_2}^2) \over (m_{\tilde{t}_2}^2
- m_{\tilde{t}_1}^2 )^2}
+ {3 m_t^2 \lambda A_t \cot \beta \cos \theta
\over 16 \pi^2 x}
f(m_{\tilde{t}_1}^2, \ m_{\tilde{t}_2}^2)  \ , \cr
& & \cr
\delta M_{12}^2 & = & {3 \over 8 \pi^2}
\left \{ {m_t^2 \lambda x \Delta_1 \over v \sin \beta}
+ {\cos \beta \Delta \over 2 v} \right \}
\left \{
{m_t^2 A_t \Delta_2 \over v \sin \beta}
- {\sin \beta \Delta \over 2 v} \right \}
{g(m_{\tilde{t}_1}^2, \ m_{\tilde{t}_2}^2)
\over (m_{\tilde{t}_2}^2 - m_{\tilde{t}_1}^2)^2}  \cr
& & \cr
& & \mbox{} + {3 \over 32 \pi^2 v^2}
\left \{
\left ({4 m_W^2 \over 3} - {5 m_Z^2 \over 6} \right)^2 \sin 2 \beta
- {2 m_t^2 \lambda x A_t \cos \theta \over \sin^2 \beta} \right \}
f(m_{\tilde{t}_1}^2, \ m_{\tilde{t}_2}^2)  \cr
& & \cr
& &\mbox{} + {3 \sin \beta \over 32 \pi^2 v}
\left ({4 m_t^2 \over \sin^2 \beta} - m_Z^2 \right)
\left \{
{m_t^2 \lambda x \Delta_1 \over v \sin \beta}
+ {\cos \beta \Delta \over 2 v} \right \}
{\displaystyle \log (m_{\tilde{t}_2}^2 / m_{\tilde{t}_1}^2)
 \over (m_{\tilde{t}_2}^2 - m_{\tilde{t}_1}^2)} \cr
& & \cr
& &\mbox{} + {3 m_Z^2 \cos \beta \over 32 \pi^2 v} \left \{
{m_t^2 A_t \Delta_2 \over v \sin \beta}
- {\sin \beta \Delta \over 2 v} \right \}
{\displaystyle \log (m_{\tilde{t}_2}^2 / m_{\tilde{t}_1}^2)
 \over (m_{\tilde{t}_2}^2 - m_{\tilde{t}_1}^2)} \cr
& & \cr
& &\mbox{} + {3 m_Z^2 \sin 2 \beta \over 256 \pi^2 v^2}
\left ({4 m_t^2 \over \sin^2 \beta} - m_Z^2 \right)
\log {m_{\tilde{t}_2}^2 m_{\tilde{t}_1}^2 \over \Lambda^4} \ , \cr
& & \cr
\delta M_{13}^2 & = & {3 \over 8 \pi^2}
\left \{ {m_t^2 \lambda x \Delta_1 \over v \sin \beta}
+ {\cos \beta \Delta \over 2 v} \right \}
{m_t^2 \lambda \cot \beta \Delta_1
\over (m_{\tilde{t}_2}^2 - m_{\tilde{t}_1}^2)^2 }
g(m_{\tilde{t}_1}^2, \ m_{\tilde{t}_2}^2) \cr
& & \cr
& &\mbox{} - {3 m_t^2 \lambda \over 16 \pi^2 v \sin \beta}
(A_t \cos \theta + 2 \lambda x \cot \beta)
f(m_{\tilde{t}_1}^2, \ m_{\tilde{t}_2}^2)  \cr
& & \cr
& &\mbox{} + {3 m_Z^2 m_t^2 \lambda \cos \beta \cot \beta \over 32 \pi^2 v}
\left ({\Delta_1 \over m_{\tilde{t}_2}^2 - m_{\tilde{t}_1}^2} \right )
\log {m_{\tilde{t}_2}^2 \over m_{\tilde{t}_1}^2}
\ , \cr
& & \cr
\delta M_{14}^2 & = & \mbox{} - {3 m_t^2 \lambda x A_t \sin \theta
\over 8 \pi^2 v \sin^2 \beta}
\left \{{m_t^2 \lambda x \Delta_1 \over v \sin \beta}
+ {\cos \beta \Delta \over 2 v} \right \}
{g(m_{\tilde{t}_1}^2, \ m_{\tilde{t}_2}^2) \over
(m_{\tilde{t}_2}^2 - m_{\tilde{t}_1}^2)^2 }  \cr
& & \cr
& &\mbox{}
+ {3 m_t^2 \lambda x A_t \cot \beta \sin \theta \over 16 \pi^2 v^2 \sin \beta}
\left \{ f(m_{\tilde{t}_1}^2, \ m_{\tilde{t}_2}^2)
- {m_Z^2 \over 2 (m_{\tilde{t}_2}^2 - m_{\tilde{t}_1}^2)}
\log {m_{\tilde{t}_2}^2 \over m_{\tilde{t}_1}^2 } \right \}
\ , \cr
& & \cr
\delta M_{15}^2 & = & \mbox{} - {3 m_t^2 \lambda A_t \cot \beta \sin \theta
\over 8 \pi^2}
\left \{{m_t^2 \lambda x \Delta_1 \over v \sin \beta}
+ {\cos \beta \Delta \over 2 v} \right \}
{g(m_{\tilde{t}_1}^2, \ m_{\tilde{t}_2}^2) \over
(m_{\tilde{t}_2}^2 - m_{\tilde{t}_1}^2)^2 }  \cr
& & \cr
& &\mbox{}
+ {3 m_t^2 \lambda A_t \cot \beta \sin \theta \over 16 \pi^2 v \sin \beta}
\left \{ f(m_{\tilde{t}_1}^2, \ m_{\tilde{t}_2}^2)
- {m_Z^2 \over 2 (m_{\tilde{t}_2}^2 - m_{\tilde{t}_1}^2)}
\log {m_{\tilde{t}_2}^2 \over m_{\tilde{t}_1}^2} \right \}
\ , \cr
& & \cr
\delta M_{23}^2 & = & {3 \over 8 \pi^2}
\left \{ {m_t^2 A_t \Delta_2 \over v \sin \beta}
 - {\sin \beta \Delta \over 2 v} \right \}
{m_t^2 \lambda \cot \beta \Delta_1
\over (m_{\tilde{t}_2}^2 - m_{\tilde{t}_1}^2)^2}
g(m_{\tilde{t}_1}^2, \ m_{\tilde{t}_2}^2) \cr
& & \cr
& &\mbox{}
- {3 m_t^2 \lambda A_t \cot \beta \cos \theta \over 16 \pi^2 v \sin \beta}
f(m_{\tilde{t}_1}^2, \ m_{\tilde{t}_2}^2) \cr
& &\mbox{} + {3 m_t^2 \lambda \cos \beta \over 32 \pi^2 v}
\left ({4 m_t^2 \over \sin^2 \beta} - m_Z^2 \right)
{\left (\Delta_1 \over m_{\tilde{t}_2}^2 - m_{\tilde{t}_1}^2 \right) }
\log {m_{\tilde{t}_2}^2  \over m_{\tilde{t}_1}^2} , \cr
& & \cr
\delta M_{24}^2 & = & \mbox{} - {3 m_t^2 \lambda x A_t \sin \theta
\over 8 \pi^2 v \sin^2 \beta}
\left \{{m_t^2 A_t \Delta_2 \over v \sin \beta}
- {\sin \beta \Delta \over 2 v} \right \}
{g(m_{\tilde{t}_1}^2, \ m_{\tilde{t}_2}^2) \over
(m_{\tilde{t}_2}^2 - m_{\tilde{t}_1}^2)^2 }  \cr
& & \cr
& &\mbox{}
+ {3 m_t^2 \lambda x A_t \sin \theta \over 16 \pi^2 v^2 \sin \beta}
\left \{ f(m_{\tilde{t}_1}^2, \ m_{\tilde{t}_2}^2)
+ {m_Z^2 \over 2 (m_{\tilde{t}_2}^2 - m_{\tilde{t}_1}^2)}
\log {m_{\tilde{t}_2}^2 \over m_{\tilde{t}_1}^2} \right \}
\ , \cr
& & \cr
\delta M_{25}^2 & = & \mbox{} - {3 m_t^2 \lambda A_t \cot \beta \sin \theta
\over 8 \pi^2}
\left \{{m_t^2 A_t \Delta_2 \over v \sin \beta}
- {\sin \beta \Delta \over 2 v} \right \}
{g(m_{\tilde{t}_1}^2, \ m_{\tilde{t}_2}^2) \over
(m_{\tilde{t}_2}^2 - m_{\tilde{t}_1}^2)^2 }  \cr
& & \cr
& &\mbox{}
+ {3 m_t^2 \lambda A_t \cot \beta \sin \theta \over 16 \pi^2 v \sin \beta}
\left \{ f(m_{\tilde{t}_1}^2, \ m_{\tilde{t}_2}^2)
+ {m_Z^2 \over 2 (m_{\tilde{t}_2}^2 - m_{\tilde{t}_1}^2)}
\log {m_{\tilde{t}_2}^2 \over m_{\tilde{t}_1}^2} \right \}
\ , \cr
& & \cr
\delta M_{34}^2 & = & \mbox{} - {3 m_t^4 \lambda^2 x A_t \cot \beta \sin \theta
\over 8 \pi^2 v \sin^2 \beta}
{\Delta_1 g(m_{\tilde{t}_1}^2, \ m_{\tilde{t}_2}^2)
\over (m_{\tilde{t}_2}^2 - m_{\tilde{t}_1}^2)^2 }
+ {3 m_t^2 \lambda A_t \sin \theta \over 16 \pi^2 v \sin^2 \beta}
f(m_{\tilde{t}_1}^2, \ m_{\tilde{t}_2}^2)  \ , \cr
& & \cr
\delta M_{35}^2 & = & \mbox{} - {3 m_t^4 \lambda^2 A_t \cot^2 \beta \sin \theta
\over 8 \pi^2}
{\Delta_1 g(m_{\tilde{t}_1}^2, \ m_{\tilde{t}_2}^2)
\over (m_{\tilde{t}_2}^2 - m_{\tilde{t}_1}^2)^2 } \ , \cr
& & \cr
\delta M_{45}^2 & = & \mbox{} {3 m_t^4 \lambda^2 x A_t^2 \cot \beta
\sin^2 \theta \over 8 \pi^2 v \sin^2 \beta}
{g(m_{\tilde{t}_1}^2, \ m_{\tilde{t}_2}^2)
\over (m_{\tilde{t}_2}^2 - m_{\tilde{t}_1}^2)^2 }
+ {3 m_t^2 \lambda A_t \cos \theta \over 16 \pi^2 v \sin^2 \beta}
f(m_{\tilde{t}_1}^2, \ m_{\tilde{t}_2}^2)  \ ,
\end{eqnarray}
with
\begin{eqnarray}
        \Delta_1 & = & A_t \cos \theta + \lambda x \cot \beta  \  , \cr
        \Delta_2 & = & A_t + \lambda x \cot \beta \cos \theta \ , \cr
        \Delta & = & \left \{(m_Q^2 - m_T^2)
+ \left ( {4 m_W^2 \over 3} - {5 m_Z^2 \over 6} \right) \cos 2 \beta
\right \} \left ( {4 m_W^2 \over 3} - {5 m_Z^2 \over 6} \right) \ ,
\end{eqnarray}
and
\[
        g(m_1^2,m_2^2) = {m_2^2 + m_1^2 \over m_1^2 - m_2^2}
        \log {m_2^2 \over m_1^2} + 2 \ .
\]
Here, in the limit of $\sin \theta = 0$, some of the elements of $\delta M^2$
vanish. In this limit, the radiatively-corrected mass matrix of the neutral
Higgs bosons reduces to the one obtained without the spontaneous violation of
the CP symmetry [7].

We also remark that the spontaneous violation of the CP symmetry
forbids a non-zero CP phase in our scenario at the 1-loop level
if $\theta=\delta$ at the tree level.
This is exactly what the Georgi-Pais theorem [13] says: The radiative
CP violation can be realized when at the tree level there
exist massless Higgs bosons other than Goldstone bosons.
That is, spontaneous CP violation does not occur in the NMSSM Higgs sector
through only radiative corrections because at the tree level there is no
pseudo-Goldstone boson.

One can notice that the scalar-pseudoscalar mixing elements
of the radiatively corrected mass matrix $\delta M^2$ of
the neutral Higgs bosons are nonzero, as far as $\theta$ is nonzero,
assuming the non-degeneracy of the left-handed and the right-handed
scalar-top quark masses.
The magnitudes of these elements are proportion to $\sin \theta$.
The scalar-pseudoscalar mixings between two Higgs doublets generated by
the radiative corrections would not occur if the degenracy of the
left-handed and the right-handed scalar-top quark masses in the 1-loop
effective potential is assumed.

The five physical neutral Higgs bosons are defined as the mass
eigenstates, obtained by
diagonalizing the mass matrix at 1-loop level, by the help of
an orthogonal transformation matrix. The elements of this orthogonal
transformation matrix determine the couplings of the physical
neutral Higgs bosons to the other states in the model.
Let us denote the physical five neutral Higgs bosons
as $h_i$ ($i$ = 1,2,3,4,5). We take the mass eigenvalues in increasing
order of the mass eigenstates $h_i$.

Constraints of the NMSSM parameter space arise from searches for
the Higgs bosons at the LEP1.
In our numerical analysis, we have used the following experimental
constraints  from LEP1.
The fact that two Higgs bosons have not been produced in the decay
of $Z$ gives the condition of $m_{h_1} + m_{h_2} > m_Z$.
In the case of $m_{h_1} + m_{h_2} < m_Z$, the decay $Z \rightarrow h_1 h_2$
is kinematically allowed and the branching ratio $B ( Z \rightarrow h_1 h_2)$
should be smaller than 10$^{- 7}$.
For two Higgs bosons $(h_1, \ h_2)$, both $B ( Z \rightarrow h_1 l^+ l^-)$
and $B ( Z \rightarrow h_2 l^+ l^-)$ should be smaller than 1.3 $\times$
10$^{- 7}$.
In our numerical analyses, the renormalization scale and
the mass of the top-quark [14] is fixed as 1000 GeV and 175 GeV, respectively.
The upper bounds on $\lambda$ and $k$ are given as 0.87 and 0.63,
respectively, by the renormalization group analysis of the NMSSM [15].
The phase $\delta$ and coupling costant $k$ is determined by two
minimum conditions (Eq. 8) of the Higgs potential.
Assuming the same soft SUSY breaking scalar-quark masses ($m_Q$ = $m_T$),
we can determine they by the relation,
$\Delta m_{\tilde t}^2/m_{\tilde t}^2 \sim 1/30$.
Here the scalar-top quark masses are constrained to be very heavy for later
convenience.

Within the context of our model, we plot the masses of the
neutral Higgs bosons by the Monte Carlo method using the above formulae
for reasonable regions of the parameter space. Fig. 1 shows
at the 1-loop level the three lighter neutral Higgs boson masses
$m_{h_1}$, $m_{h_2}$, and $m_{h_3}$ as functions of $A_t$,
for $0 < \theta <  2\pi$, $1 < \tan\beta < 10$, $0 < \lambda < 0.87$,
and $0 < A_{\lambda}, - A_k, x <$ 1000 GeV.
One can see in Fig. 1a that many points are
excluded for the range of $m_{h_1} < m_Z$ by the negative
experimental results at LEP1; but the LEP1 data do not completely exclude the
existence of a massless neutral Higgs boson of the model at the
1-loop level.

Our numerical result does not seem to be compatible with that of
recent research [10].
It was recently pointed out that at the 1-loop
level a wide region of the NMSSM parameter space is excluded for
spontaneous CP violation. Especially the range of $\tan \beta>$ 1
is completely ruled out in this scenario and then in this analysis
$m_{h_1}$ is relatively small ( $\sim$ 35 GeV) for most of the
parameter space. Here Haba at al. [10] did not consider the range of
$A_k <$ 0 in their analysis. The sign of $A_k$ values need not be
positive when CP is spontaneously violated in the Higgs sector.

The range of $A_k <$ 0 plays a important role in allowing spontaneous
CP violation in the NMSSM because of in the minimum equation the
sign of $k$ is negative for a reasonable parameter space.
Thus the range of $\tan \beta >$ 1 is allowed for spontaneous CP violation
in the Higgs sector of the NMSSM in our analysis.
In the parameter space of Fig. 1 but the range of $A_k >$ 0, spontaneous CP violation
is not allowded in the NMSSM even though radiative corrections are
included. Figs. 1b, 1c display the allowed ranges of 60 $< m_{h_2}
<$ 245 GeV for $m_{h_2}$ and of 100 $< m_{h_2} <$ 525 GeV for
$m_{h_3}$.
In these figures the upper and lower bounds on the
neutral Higgs mass boson masses are theoretical and experimental
bounds, respectively.
We also calculated $m_{h_1}$ for the same parameter space as that of Fig. 1
without spontaneous CP violation in the NMSSM. The upper bound on
$m_{h_1}$ is decreased by spontaneous CP violation in the Higgs
sector.
\section{$K$-$\bar{K}$ MIXING IN NMSSM}
\hspace*{6.mm}
We now explore their phenomenological implications for spontaneous
CP-violating effects in $K$-${\bar K}$ mixing.
In the SM, the $K$-${\bar K}$ mixing arises from the $W$-exchange box
diagram giving rise to the $\Delta S$ = 2 operator at the 1-loop
level.
In the NMSSM, there are dominant sources for CP violation in
the $K$-${\bar K}$
system through box diagrams involving superpartners (superbox).
In the superbox diagrams, the gauge fermion couplings are
described by the super-CKM unitary matrices which diagonalize the
scalar quark mass matrix [16].
The super-CKM matrix is real in the spontaneous CP violation
scenario that we concentrate on.

After electroweak symmetry breaking takes place, the gauginos and
Higgsinos with the same spin combine to form mass eigenstates,
charginos and neutralinos.
The complex mixing between gauginos and Higgsinos induce a complex
chargino or neutralino propagator [17].
In the case, the superbox diagrams contain couplings among quark
scalar-quark, and Higgsino
so that their contribution to the imaginary part of $K$-${\bar K}$
mixing is suppressed by a factor of $m_q/m_W$.
If the superbox diagrams contain the weak phase in a scalar quark
propagator, the superbox digrams receive a suppression factor of
$m_q/m_{\tilde q}$.
Thus, the dominant contributions to the imaginary part
of $K$-${\bar K}$ mixing
come from those including the left-right scalar-top quark mixing.
The superbox diagrams for these contributions are displayed in
other papers [17, 18].
Assuming that scalar-top quarks are very heavy $m_{\tilde t} \gg m_{\tilde W}$
and the upper bound on $\Delta m_{\tilde t}^2/m_{\tilde t}^2$ is
saturated as $1/30$, the ratio of the imaginary and real parts of
the $K$-${\bar K}$ mixing can be described by the quantity [18]
\begin{equation}
      3 \left ( {m_t \over v \sin \beta} \right)^2
        {v m_{\rm LR}^2 V_{13} z \sin \phi \over m_{\tilde W} m_{\tilde
        t}^2}
      \left( {\Delta m_{\tilde t}^2 \over m_{\tilde t}^2}
      \right)^{- 1} \ ,
\end{equation}
where $z = 0.5$ is a partial cancellation factor, $m_{\rm LR}$ is
the left-right scalar-top quark mixing, $m_{\tilde t}$ is the
average of the scalar-top quark masses, $V_{13} = 10 ^{- 2}$ is
a element of the super-CKM matrix for the scalar-quark, and $m_{\tilde W}$
is the superpartner (wino) of SU(2).
The CP-violating phase $\phi$ is expressed as a function of two
phases ($\theta$, \ $\delta$) of VEV's.

We impose the experimental constraints at LEP1 on the NMSSM parameter space
and then estimate numerically CP-violating effects in $K$-${\bar K}$ mixing by
using the parameters obtained in the context of spontaneous CP-violating
scenario.
The experimental upper bound on the ratio of $K$-${\bar K}$ mixing is given by
6 $\times$ 10$^{- 3}$ [16].
This bound impose a constraint on the weak CP phase.
The wino is not exist as a mass eigenstate.
Thus we fix as $m_{\tilde W}$ = 100 GeV for heavy
scalar-top quarks.
We use Eq. 11 for the ratio of the $K-{\bar K}$ mixing and plot in
Fig. 2a the lower bound on the weak phase $\phi$  as a function of $A_t$, for
0 $< \tan \beta <$ 10, 0 $< \theta <$ 2$\pi$, 0 $< \lambda<$ 0.87, and
0 $< A_{\lambda}, - A_k, x <$ 1000 GeV.
One can notice that the lower bound on $\phi$ becomes small as the value of
$A_t$ decrease.
Fig. 2b shows the lower bound on the weak phase $\phi$ as the same
parameter space of as that of Fig. 1a but as a function of $\tan \beta$.
In Fig. 2b the lower bound on $\phi$ increases as the values of
$\tan \beta$ decrease.
\section{CONCLUSIONS}
\hspace*{6.mm}
Spontaneous CP violation is investigated in the Higgs sector of the
NMSSM.
In previous analyses [9], it has been shown that the spontaneous
violation of the CP symmetry can be realized in the NMSSM Higgs sector
by radiative corrections coming from the degenerate scalar-top quark
masses. In this case, however, the presence of the CP violating vacuum
requires a very light Higgs boson in the model.
By considering the negative results in the Higgs search at LEP1,
the possibility of the spontaneous violation of the CP symmetry is
almost completely excluded in the Higgs sector of the NMSSM,
if the scalar-top quark masses are degenerate.
The numerical analysis also have shown [10] that the situation is
hardly improved when the radiative corrections including the mass
spliting effect between the scalar-top quarks are taken into account.

Here we reanalyze the possibility of spontaneous CP violation in the
Higgs sector of the NMSSM with the non-degeneracy of the scalar-top quark
masses in the 1-loop effective potential.
In this analysis the Higgs potential contains radiative
corrections due to the top-quark and scalar-top quark
contributions. In the NMSSM the mass matrix of the neutral Higgs
boson is analytically derived in spontaneous CP violation scenario.
In the range of $A_k >$ 0 CP violation is not realized even though
radiative corrections due to the these contributions are included
in the Higgs sector.
On the contrary, CP violation scenario is viable in the range of
$A_k <$ 0.
Assuming both scalar-top quarks are much heavier than the wino and
scalar-top quarks are maximally degenerated in their masses,
the neutral Higgs boson masses are numerically calculated.
The three neutral Higgs bosons are relatively light.
The upper bound on the lightest neutral Higgs boson mass is about
140 GeV under our choice for the parameter space.
This upper bound increases if spontaneous CP violation does not occur
in the Higgs sector.
In the 1-loop parameter space of the NMSSM, the lower bound on the
phase $\phi$ is calculated by using the experimental result for the
ratio of $K$-${\bar K}$ mixing.
The lower bound on the phase $\phi$ increase as $\tan \beta$
values decrease or as $A_t$ values increase.
In the NMSSM Higgs sector, a possibility of spontaneous CP violation
scenario can not be completely ruled out by the Higgs search at LEP.

\vskip 0.3 in

\noindent
{\large {\bf ACKNOWLEDGMENTS}}

This work was supported by the Korea Science and Engineering Foundation (KOSEF)
through the Center for Theoretical Physics (CTP) at Seoul National University
(SNU).
\vspace{-1cm}

\vfil\eject
{\bf Figure Captions}
\vskip 0.3 in
\noindent
Fig. 1 : \ (a) $m_{h_1}$, (b) $m_{h_2}$, and (c) $m_{h_3}$
as a function of $A_t$, for 0 $< \theta <$
2$\pi$, 1 $< \tan \beta <$ 10, 0 $< \lambda<$ 0.87, and 0
$< A_{\lambda}, - A_k, x <$ 1000 GeV.
\vskip 0.2 in
\noindent
Fig. 2 : \  The lower bound on $\phi$ as a function of (a) $A_t$
and (b) $\tan \beta$

\newpage
\begin{figure}[t]
\epsfxsize=15cm
\hspace*{0.cm}
\epsffile{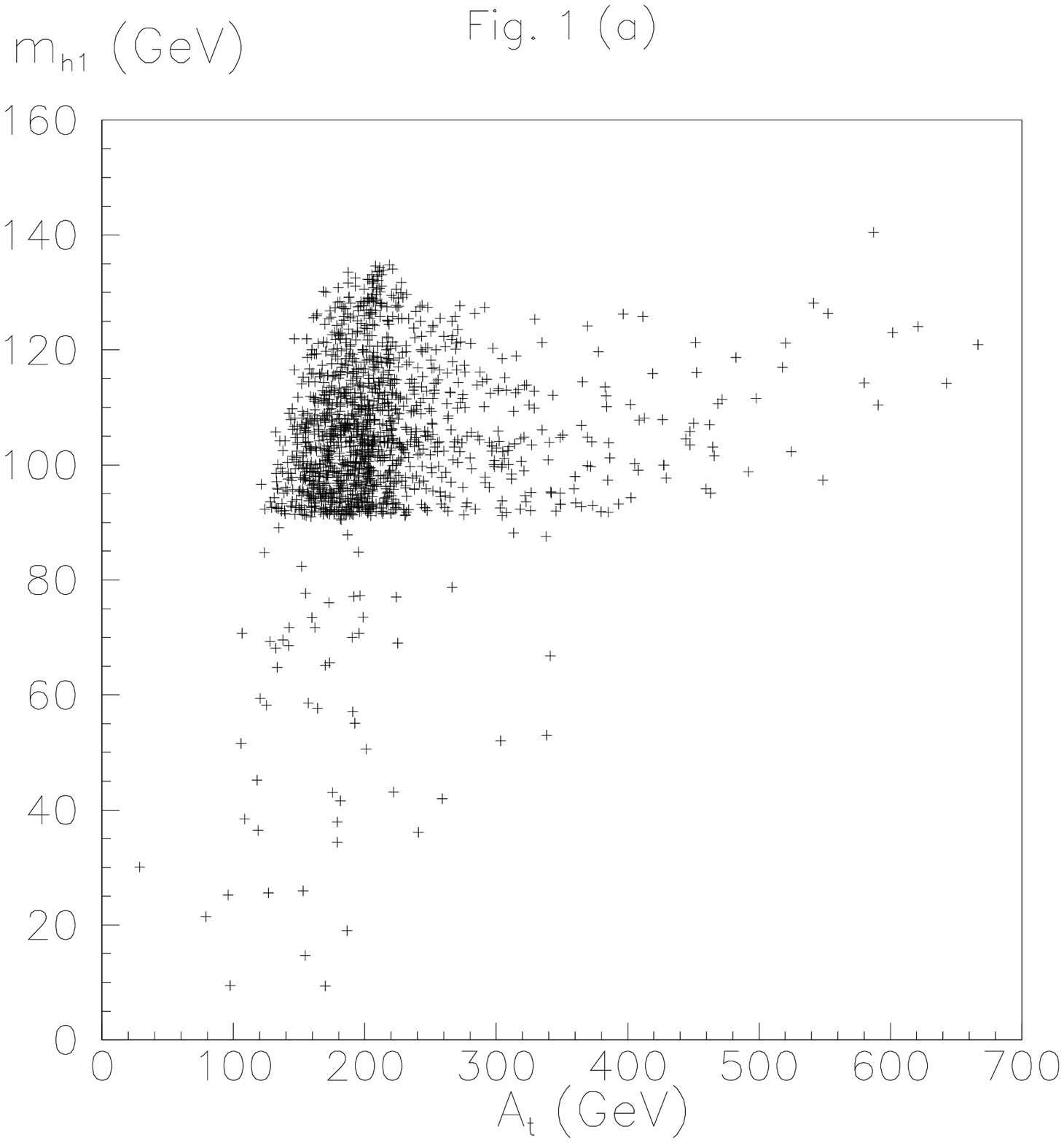}
\caption[plot]{fig1a}
\end{figure}
\begin{figure}[t]
\epsfxsize=15cm
\hspace*{0.cm}
\epsffile{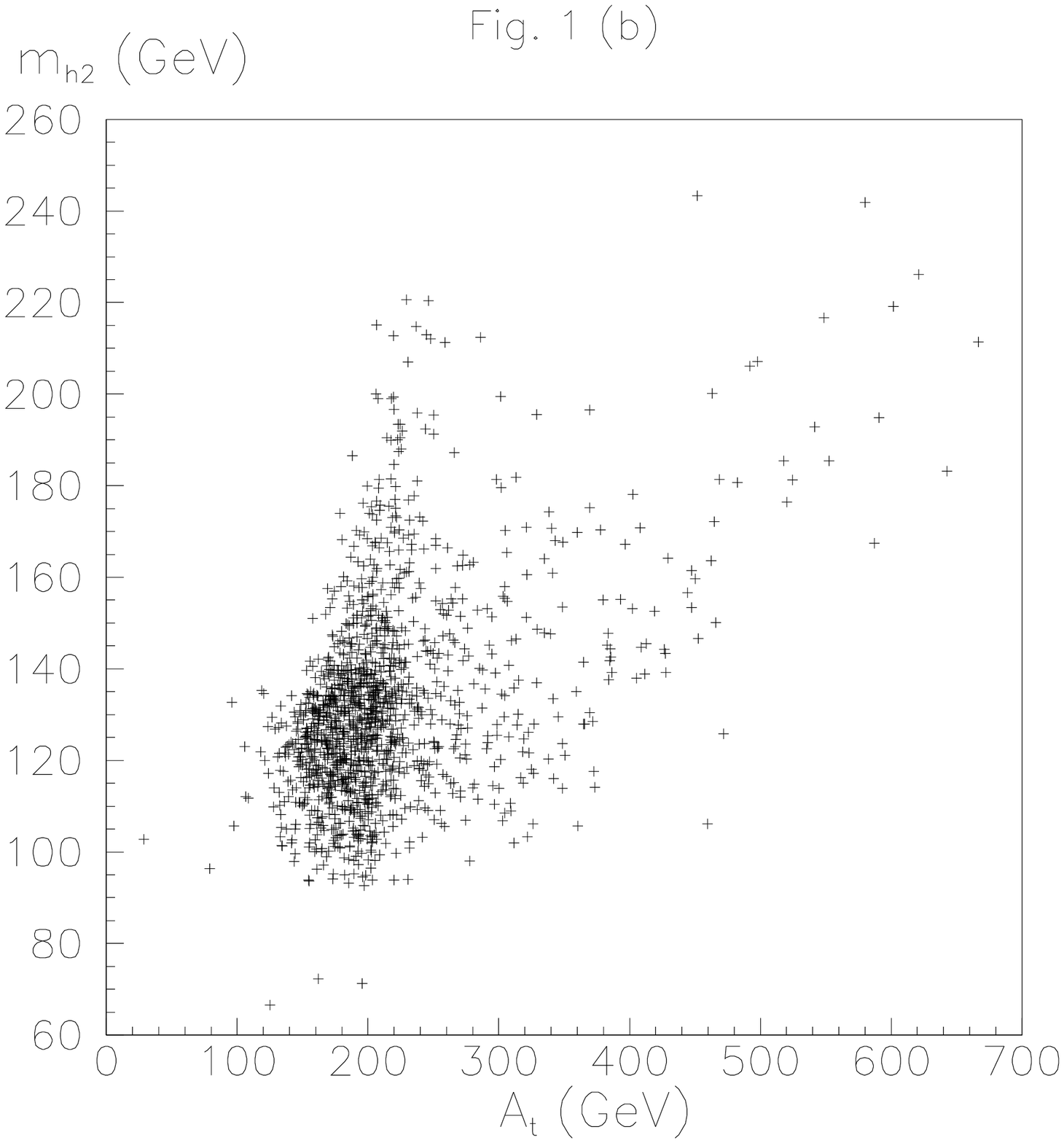}
\caption[plot]{fig1b}
\end{figure}
\begin{figure}[t]
\epsfxsize=15cm
\hspace*{0.cm}
\epsffile{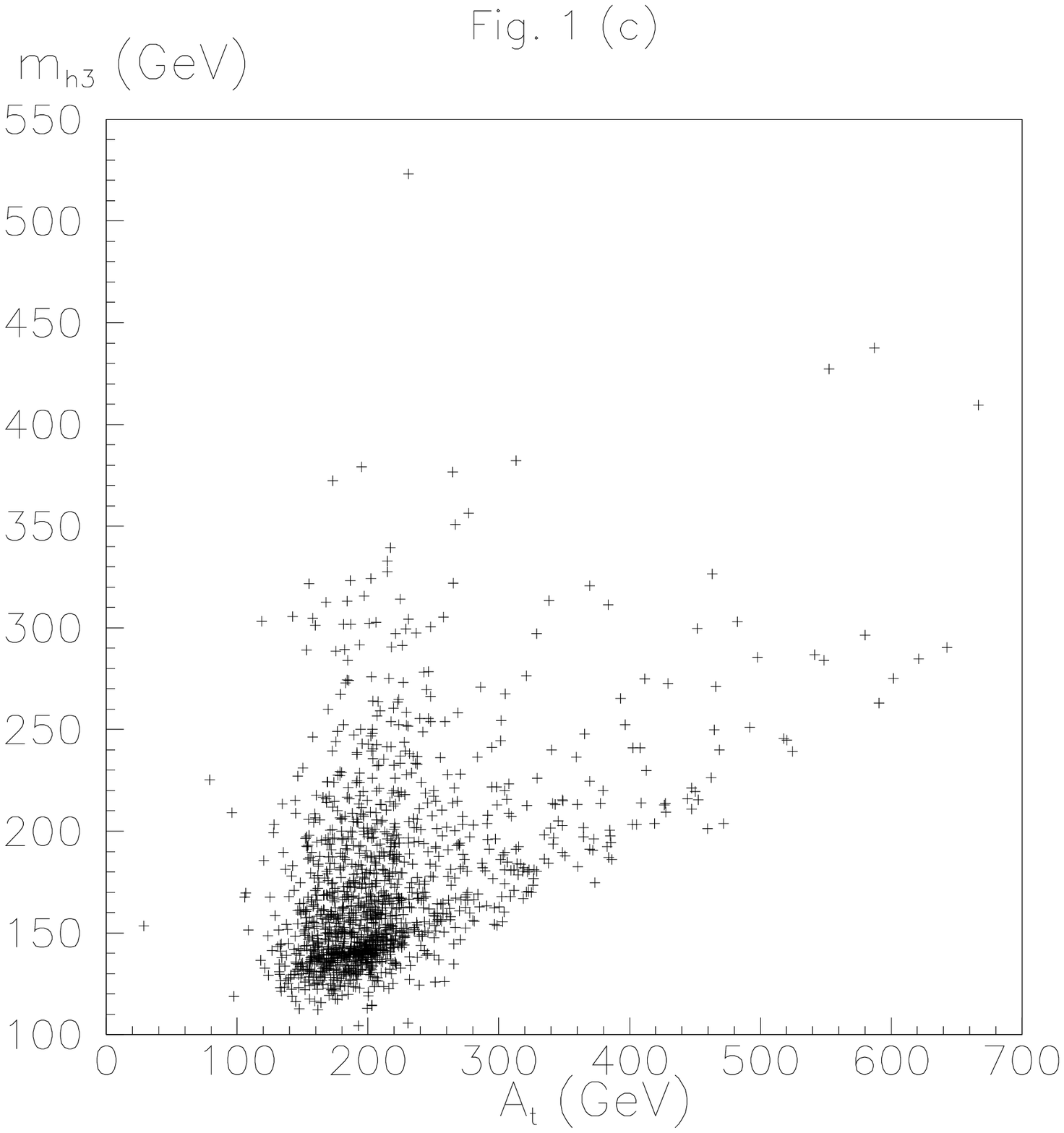}
\caption[plot]{fig1c}
\end{figure}
\begin{figure}[t]
\epsfxsize=15cm
\hspace*{0.cm}
\epsffile{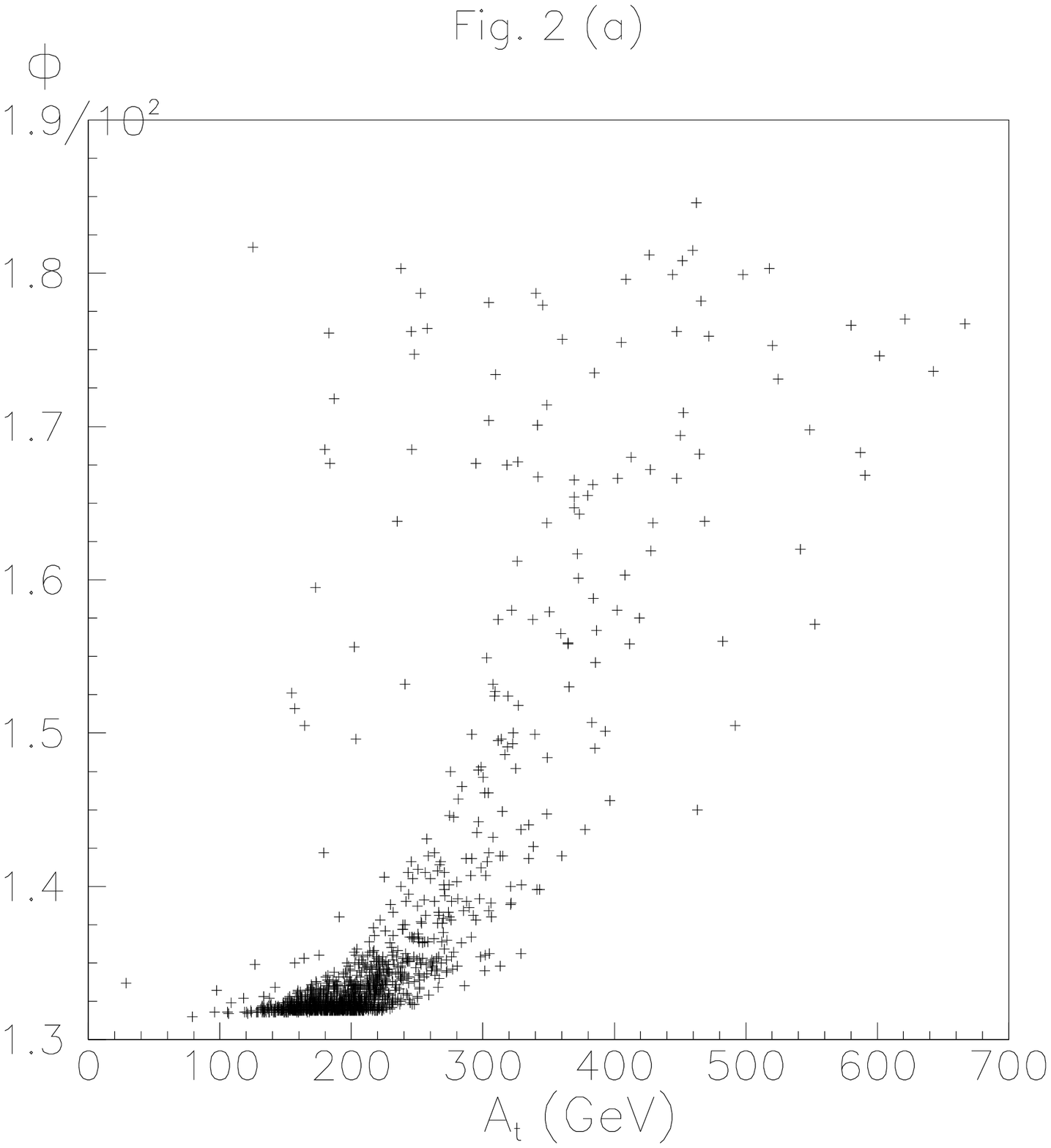}
\caption[plot]{fig2a}
\end{figure}
\begin{figure}[t]
\epsfxsize=15cm
\hspace*{0.cm}
\epsffile{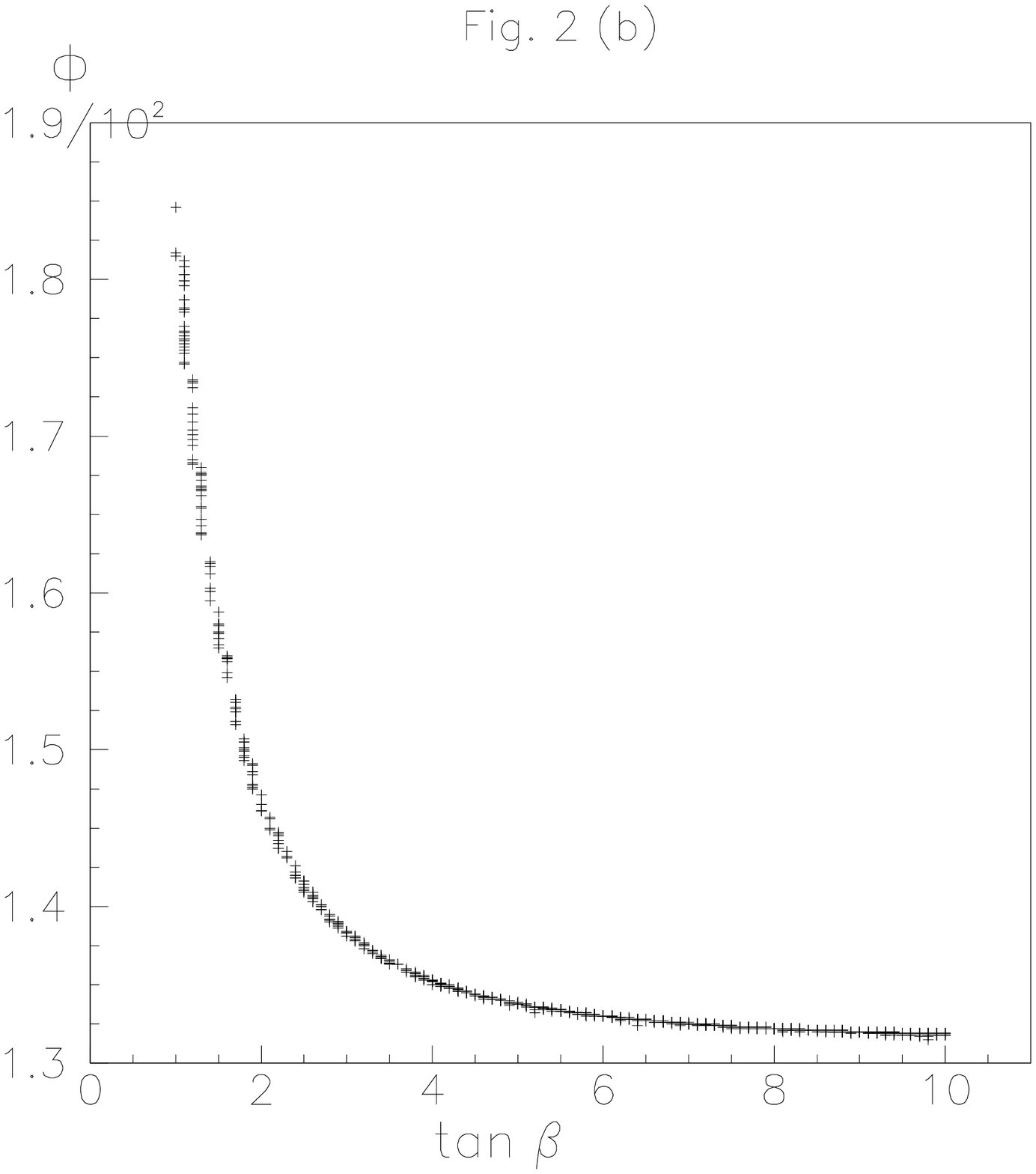}
\caption[plot]{fig2b}
\end{figure}

\begin{thebibliography}{99}
\bibitem{1} T.D. Lee and C.N. Yang, Phys. Rev. {\bf 104}, 254 (1956);
J.H. Christenson, J.W. Cronin, V.L. Fitch and R. Turlay,
Phys. Rev. Lett. {\bf 13}, 138 (1964).
\bibitem{2} S. Glashow, Nucl. Phys. {\bf 22}, 475 (1961); S. Weinberg,
Phys. Rev. Lett. {\bf 19}, 1264 (1967); A. Salam, in
{\it Proc. 8th Nobel Symposium, 367}, edited by N. Svartholm
(Almquist and Wiksell, Stockholm, 1968).
\bibitem{3}
N. Cabibbo, Phys. Rev. Rett. {\bf 10}, 531 (1963).
M. Kobayashi and T. Maskawa, Prog. Theor. Phys.
{\bf 49} 652 (1973).
\bibitem{4} H.P. Nilles, Phys. Rep. {\bf 110}, 1 (1984)
;J.F. Gunion, H.E. Haber, G.L. Kane, and S. Dawson,
{\it The Higgs Hunters' Guide} (Addison-Wesley Pub. Co., Redwood
City, CA, USA, 1990).
\bibitem{5}H. E. Haber, R. Hempfling, and Y. Nir, Phys. Rev.
D {\bf 46} 3015 (1992);
M. A. Diaz and H. E. Haber, Phys. Rev. D {\bf 45}, 4246 (1992);
H. E. Haber, R. Hempfling, Phys. Rev. Lett. {\bf 66}, 1815 (1991);
A. Brignole, J. Ellis, G. Ridolfi, and F. Zwirner,
Phys. Lett. B {\bf 271}, 123 (1991).
\bibitem{6} O. Lebedev, Eur. Phys. J. C {\bf 4}, 363 (1998);
N. Haba, Phys. Lett. B {\bf 398}, 305 (1997);
A. Pomarol, Phys. Lett. B {\bf 287}, 331 (1992);
N. Maekawa, Phys. Lett. B {\bf 282}, 387 (1992).
\bibitem{7}
S.W. Ham, S.K. Oh, and B.R. Kim, Phys. Lett.
B {\bf 414}, 305 (1997);
S.W. Ham, S.K. Oh, and B.R. Kim, J. Phys.
G {\bf 22}, 1575 (1996);
F. Franke and H. Fraas, Phys. Lett. B {\bf 353}, 234
(1995);
S.F. King and P.L. White, Phys. Rev.
D {\bf 52}, 4183 (1995);
C.M. Kim, H.J. Jahng, S.K. Oh, and B.R. Kim,
J. Korean Phys. Soc. {\bf 28} 450 (1995);
T. Elliott, S.F. King, and P.L. White, Phys. Rev.
D {\bf 49}, 2435 (1994);
T. Elliott, S.F. King, and P.L. White, Phys. Lett.
B {\bf 314}, 56 (1993);
T. Elliott, S.F. King, and P.L. White, Phys. Lett.
B {\bf 305}, 71 (1993);
U. Ellwanger and M.L. Linder, Phys. Lett.
B {\bf 301}, 365 (1993);
P.N. Pandita, Phys. Lett. B {\bf 318}, 338 (1993);
U. Ellwanger, Phys. Lett. B {\bf 303}, 271 (1993);
U. Ellwanger, M.R. de Traubenberg, and C.A. Savoy,
Phys. Lett. B {\bf 315}, 331 (1993);
U. Ellwanger and M.R. de Traubenberg, Z. Phys.
C {\bf 53}, 521 (1992);
P. Bin\'{e}truy and C.A. Savoy, Phys. Lett.
B {\bf 277}, 453 (1992).
\bibitem{8} J.C. Rom${\tilde {\rm a}}$o, Phys. Lett. B {\bf 173},
309.
\bibitem{9} K.S. Babu and S.M. Barr, Phys. Rev D {\bf 49}, R2156
(1994).
\bibitem{10} N. Haba, M. Matsuda, and M. Tanimoto, Phys. Rev D {\bf 54},
6928 (1996).
\bibitem{11} S. Coleman and E. Weinberg, Phys. Rev. D {\bf 7}, 1888
(1973).
\bibitem{12} N. Haba, Prog. Theor. Phys. {\bf 97}, 301 (1997);
M. Matsuda and M. Tanimoto, Phys. Rev D {\bf 52}, 3100 (1995).
\bibitem{13} H. Georgi abd A. Pais, Phys. Pev. D {\bf 10}, 1246
(1974).
\bibitem{14} CDF Collaboration, F. Abe {\it et al.}, Phys. Rev. Lett.
{\bf 80}, 2767 (1998); D0 Collaboration, S. Abachi {\it et al.}, Phys.
Rev. Lett. {\bf 79}, 1197 (1997).
\bibitem{15} J. Ellis, J.F. Gunion, H.E. Haber, L. Roszkowski,
and F. Zwirner, Phys. Rev. D {\bf 39}, 844 (1989).
\bibitem{16} A. Pomarol, Phys. Lett. B {\bf 287}, 331 (1992).
\bibitem{17} J. Ellis and D.V. Nanopoulos, Phys. Lett. B {\bf
110}, 44 (1982).
\bibitem{18} O. Lebedev, Phys. Lett. B {\bf 452}, 294 (1999);
Ph. D. Thesis, Virginia Tech, 1998.
\end{thebibliography}
\end{document}